\documentclass[
aps,%
11pt,%
final,%
notitlepage,%
oneside,%
twocolumn,%
nobibnotes,%
nofootinbib,%
superscriptaddress,%
noshowpacs,%
centertags]%
{revtex4}

\newcommand{\sai}{Sternberg Astronomical Institute, Universitetskii pr. 13, Moscow, 119991 Russia}

\begin{document}
\selectlanguage{english}

\title{Ultra-luminous X-ray Sources as Supercritical Accretion Disks: Spectral Energy Distributions}

\author{\firstname{A.}~\surname{Vinokurov}}
\affiliation{\saoname}

\author{\firstname{S.}~\surname{Fabrika}}
\affiliation{\saoname}

\author{\firstname{K.}~\surname{Atapin}}
\affiliation{\saoname}
\affiliation{\sai}

\begin{abstract}

We describe a model of spectral energy distribution in supercritical accretion disks (SCAD) based on the conception by Shakura and Sunyaev. We apply this model to five ultra-luminous X-ray sources (ULXs). In this approach, the disk becomes thick at distances to the center less than the spherization radius, and the temperature dependence is  $T \propto r^{-1/2}$. In this region the disk luminosity is $L_{\rm bol} \sim L_{\rm Edd}  \ln (\dot{M} / \dot{M}_{\rm Edd})$, and strong wind arises forming a wind funnel above the disk. Outside the spherization radius, the disk is thin and its total luminosity is Eddington, $L_{\rm Edd}$. The thin disk heats the wind from below. From the inner side of the funnel the wind is heated by the supercritical disk. In this paper we do not consider Comptonization in the inner hot winds which must cover the deep supercritical disk regions. Our model is technically similar to the DISKIR model of Gierlinski et al. The models differ in disk type (standard---supercritical) and irradiation (disk---wind). We propose to distinguish between these two models in the X-ray region $\sim 0.3$--$1$~keV, where the SCAD model has a flat $\nu F_{\nu}$ spectrum, and the DISKIR model never has a flat part, as it is based on the standard $\alpha$-disk. An important difference between the models can be found in their resulting black hole masses. In application to the ULX spectra, the DISKIR model yields black hole masses of a few hundred solar masses, whereas the SCAD model produces stellar-mass black holes $\sim 10$~M$_{\sun}$.

\end{abstract}

\maketitle

\section{Introduction}

Ultra-luminous X-ray sources (ULXs) are X-ray sources with luminosities exceeding the Eddington limit for a typical stellar-mass black hole~\cite{FengSoria2011}; their luminosities are $10^{39}$--$10^{41}$~erg\,s$^{-1}$. The most popular models for ULXs involve either intermediate-mass black holes (IMBH, $10^3$--$10^4$~M$_{\sun}$) or stellar-mass black holes ($\sim10$~M$_{\sun}$) accreting at highly super-Eddington rates. Both scenarios require massive donors in a close orbit. 

IMBHs originating from low-metallicity Population III stars~\cite{Madau2001} can form binary systems due to tidal captures of single stars. However, the expected frequencies of such IMBHs are not high and do not agree with the observed ULX frequencies~\cite{Kuranov2007}. Most of the ULXs are associated with star formation regions~\cite{Swartz2009} and young star clusters~\cite{Poutanen2012}. They are not distributed throughout the galaxy as would be expected for IMBHs originating from low-metallicity Population III stars. 

IMBHs may be produced in the process of runaway merging in the cores of young stellar clusters~\cite{PortegiesZwart2004}; in this case they would stay within the clusters. Poutanen et al.~\cite{Poutanen2012} have found that the majority of the brightest X-ray sources in the Antennae galaxies are located close to very young stellar clusters, but not within them. They concluded that the sources have been ejected in the process of star cluster formation due to dynamical few-body encounters and that the ULXs are massive X-ray binaries with the progenitor masses larger than $\sim50$~M$_{\sun}$.

The model where the ULXs are supercritical accretion disks (SCADs) requires geometrical collimation of radiation into the observer's line of sight~\mbox{\cite{FabrMesch2001,Poutanen2007}}; this model explains the objects up to possibly $10^{41}$~erg\,s$^{-1}$. It was suggested the ULXs are face-on copies of SS\,433, the only known supercritical microquasar in the Galaxy~\cite{FabrObz}. The binary SS\,433 contains a stellar-mass black hole accreting in a supercritical regime at $\sim300$--$500$ Eddington mass accretion rates. 
In this system the wind forms a wide funnel along the disk axis, which collimates the disk radiation. In our recent optical spectroscopy of ULX counterparts~\cite{Fabrika2013}, we have discovered that their optical spectra are very similar to that of SS\,433.

If the ULXs are SCADs with stellar-mass black holes, one may expect to detect discrete lines or edges in their spectra because of the massive outflows in the supercritical regime. The X-ray spectra of ULXs do not show any emission or absorption features; the best quality data~\cite[e.g.,][]{Walton2012} yield an upper limit on the equivalent widths of a few tens of eV. A common property of the ULX spectra is a high-energy curvature~\cite{Stobbart2006,Gladstone2009,Caballero2010} with a downturn observed between $\sim 4$ and  $\sim 7$~keV. The curvature suggests that the ULX accretion disks are not standard. The inner parts of the disks may be covered with an outflow or with a relatively cool and optically thick corona~\mbox{\cite{Stobbart2006,Gladstone2009}}, which Comptonize the inner disk photons. 

The ULXs are frequently surrounded by nebulae~\cite{Pakull2002,Abolmasov2007}; the nebula shapes are similar to those powered by jets or weakly collimated winds. In some ULXs, radial velocity perturbations in the associated nebulae were detected~\cite{Lehmann2005}. 

Strong radiative beaming is not necessary for the ionization of the ULX nebulae. Intermediate geometrical beaming ($B \sim 3$--$5$) is enough to produce both the observed luminosities and the observed ionized nebula shapes. Besides that, not only X-rays may ionize the nebulae. In the optical study of MF\,16, the nebula surrounding NGC\,6946~ULX-1, Abolmasov et al.~\cite{Abolmasov2008} have found that UV and far UV radiation is able to produce the MF\,16 spectrum; they suggested that the ULX systems may also be ultra-luminous in UV. It has been found that at least some of the ULXs are indeed very bright UV sources~\cite{Kaaret2010,TaoHoII2012}. However, both the IMBHs and SCADs are expected to be bright UV sources. 

Several distinctive properties of ULXs may be explained in these two competitive models. In spite of much effort in the ULX studies, the main questions still remain open---what are the black hole masses in ULXs, do they have standard accretion disks, do they constitute a homogeneous class of objects. 

One may suggest that the common properties of ULXs are: a hot disk wind which Comptonizes the inner disk radiation, making a Compton tail in the \mbox{X-ray} spectra, and a very luminous UV/optical source. The ULX optical counterparts show hot blue spectra; the absolute magnitudes of the nearest and well-studied ULX star-like counterparts (not clusters) in the HST images are \mbox{${\rm M_V} \approx -6 \pm 1$~mag~\cite{Tao2011}}. 

Both the high UV/optical luminosities and the Compton tails may be reconciled with a self-irradiating accretion disk. Such a model has been suggested by~\mbox{\cite{Gierlinski2008,Gierlinski2009}}, it was originally developed for LMXBs. If the disk is flat, as is the case in the standard disk model, one finds the temperature dependence $T(r) \propto r^{-3/4}$, the same as in the disks without self-heating. The authors suggested that the disk, for some reason, is warped. In this case, one may expect a flatter dependence, $T(r) \propto r^{-1/2}$, providing strong UV and optical fluxes from the outer parts of the disk. This DISKIR model has been successfully applied in several ULXs~\cite{Grise2012,BergheaMF162012,TaoHoII2012}. 

The SCADs may potentially explain both the high UV/optical luminosities and the power-law Compton tails in ULXs. The disk wind photosphere may be a luminous source with a hot black body-like spectrum~\cite{Poutanen2007,FabrObz}. Deep inside the supercritical disk funnel, where the jet is formed and collimated, hot winds make favorable conditions for the Comptonization of the surrounding photons radiated by the funnel walls. The hard radiation is expected to be mildly collimated by the funnel. The outer walls of the funnel constitute the SCAD's wind photosphere. The jets may power and shape the nebulae associated with the ULXs as it is in SS\,433.  

The supercritical accretion regime was first described by Shakura and Sunyaev (1973)~\cite{SS73}. They introduced a ``spherization radius'' in the disk, $R_{\rm sp}$, below which the disk becomes supercritical. The supercritical disk is thick with a strong mass loss. In the supercritical region, wind inevitably appears; its velocity is close to virial, as in stellar winds. At highly supercritical accretion rates ($\dot M_0 >> \dot M_{\rm Edd}$), the extended wind photosphere ($R_{\rm ph} >> R_{\rm sp}$) conceals the wind formation region, as is observed in SS\,433. 

The recent 2D RHD simulations~\cite[e.g.,][]{Ohsuga2005,Okuda2009}, which take into account both the heat advection and photon trapping, confirm the main ideas of the Shakura--Sunyaev SCAD model. Poutanen et al.~\cite{Poutanen2007} have considered ULXs within the framework of the Shakura--Sunyaev approach; they confirmed that the main features of the ULX spectra may be formed in the extended photospheres.

In this study, we simulate the spectral energy distributions (SEDs) of SCADs based on the Shakura--Sunyaev approach. We find that this model is similar in the main features to the DISKIR model~\mbox{\cite{Gierlinski2008,Gierlinski2009}}, however, the models differ in many details. In our model, the inner parts of the supercritical disk \mbox{($R < R_{\rm sp}$)} heat the outer wind, whereas in the Gierlinski et al. model the inner regions of the accretion disk heat the outer parts of the disk. In this paper, we have not yet included the Comptonization in the model, and consider the model to be valid up to  $\lesssim1.5$~keV. We plan to introduce the Comptonization in the next paper. 

It is important to have UV data points as close to the X-ray range as possible, because the spectral shape in this region may be critical for distinguishing between the models. Photometric data in \mbox{$<2000$~\AA}, bands are available only for two ULX counterparts. 
It is also important to have simultaneous data. When the variability of the ULX counterparts is not very high in the optical range (about 10--20\%,~\cite{Tao2011}), the X-ray variability may reach a factor of a few. We have found only two ULX counterparts with simultaneous X-ray---optical data. Below we describe the observations and the SED model, and compare the observed and model SEDs.

\section{Observations and data reduction}

The catalog of ultra-luminous X-ray sources of Swartz et al. 2004~\cite{Swartz2004} provides data for more than 150 objects, however, less than 50 objects are identified with optical sources~\cite{Ptak2006}. There are considerably less thoroughly studied point sources for which the spectral energy distributions are known in a wide range of wavelengths (\cite{Tao2011}). We have selected five ULXs that have reliable optical identifications with point sources: \mbox{Holmberg\,II~X-1}~\cite{Kaaret2004}, NGC\,6946~ULX-1~\cite{Kaaret2010}, NGC\,1313~X-1~\cite{Yang2011}, NGC\,1313~X-2~\cite{Liu2007} and NGC\,5408~X-1~\cite{Lang2007}. They are bright objects located in nearby galaxies (up to 5 Mpc) that have various spectral indices in the UV-optical range.  

Table~\ref{data} lists the basic information on the studied ultra-luminous X-ray sources, including the distances to the host galaxies, the adopted values of interstellar extinction, the X-ray luminosities of the objects and the spectral indices in the UV-optical range.

\begin{table*}[th]
\setcaptionmargin{0mm} \onelinecaptionstrue
\captionstyle{normal}
\caption{Observational parameters of ULXs. The columns give: D---distance to the host galaxy in Mpc; $A_V$---total extinction, obtained from nebula spectroscopy; $A_{V_{\rm G}}$---extinction in the Galaxy; $L_X$---extinction-corrected X-ray luminosity in the $0.3$--$10$~keV range in units of $10^{39}$~erg\,s$^{-1}$; $\alpha$---UV-optical spectral index, $F_{\nu} \propto {\nu}^{-\alpha}$}
\label{data}
\medskip
\begin{tabular}{l|c|c|c|c|c}
\hline
~~~~~~~~~~Name &  \hspace{5mm}D\hspace{5mm} & $\hspace{5mm}A_V\hspace{5mm}$ & $\hspace{5mm}A_{V_{\rm G}}^{ 7}\hspace{5mm}$ & $\hspace{5mm}L_X\hspace{5mm}$ &  $\hspace{5mm}\alpha\hspace{5mm}$ \\
\hline
Holmberg\,II~X-1 & ~~$3.39^{ 1}$ & $\	0.23^{~}$  & $0.10$ & ~~$6.0^{ 8}$  & $-1.07\pm0.05$\\
NGC\,6946~ULX-1 & $4.5^{ 2}$   & $\	1.34^{~}$  & $1.14$ & ~~$3.8^{ 8}$  & $-1.78\pm0.25$\\
NGC\,1313~X-1 & ~~$4.13^{ 3}$    & $\	...$       & $0.34$ & $15.7^{ 9}$ & $-0.81\pm0.03$\\
NGC\,1313~X-2 & ~~$4.13^{ 3}$    & $\	0.40^{ 5}$ & $0.26$ & ~~$5.4^{ 9}$  & $-1.34\pm0.04$\\
NGC\,5408~X-1 & $4.8^{ 4}$     & $\	0.25^{ 6}$ & $0.21$ & ~$10.2^{ 10}$ & $-1.32\pm0.07$\\
\hline
\end{tabular}
\begin{flushleft}
\footnotesize{$^1$Karachentsev et al. 2002~\cite{Karachentsev2002}, ~$^2$N. A. Tikhonov~\cite{Tikhonov}, ~$^3$Mendez et al. 2002~\cite{Mendez2002}}, ~$^4$Karachentsev et al. 2002~\cite{KarachentsevNGC54082002}, $^5$Grise et al. 2008~\cite{Grise2008}, $^6$Kaaret \& Corbel 2009~\cite{KaaretCorbel2009}, $^7$Schlegel et al.~\cite{Schlegel1998}, $^8$Swartz et al. 2011~\cite{Swartz2011}, $^{9}$average value based on the data from Feng \& Kaaret 2006~\cite{FengKaaret2006}, $^{10}$Grise et al. 2012~\cite{Grise2012}
\end{flushleft}
\end{table*}

\subsection{Optical data}
\label{opticdata}
We use the HST archive data. The observations were carried out with the Advanced Camera for Surveys (ACS), the Wide Field and Planetary Camera~2 (WFPC2) and the Wide Field Camera 3 (WFC3) (the last one was used only for the observations of NGC\,5408~X-1). 

For the objects NGC\,1313~X-1, NGC\,1313~X-2 and NGC\,5408~X-1 we used the photometric results from~\cite{Tao2011,Grise2012,Yang2011}. We selected quasi-simultaneous observations with a maximum set of filters. For NGC\,1313~X-1 we took the flux measurements in filters F330W, F435W, F555W and F814W taken on November 17, 2003, from~\cite{Yang2011}. For \mbox{NGC\,1313~X-2} the photometric measurements in filters F330W, F435W, F555W and F814W on November 22, 2003, were taken from~\cite{Tao2011}; for NGC\,5408~X-1 we used the results of the photometry in filters F225W, F336W, F547M, F845M, F105W and F160W on May 15, 2010, from~\cite{Grise2012}.

For Holmberg\,II~X-1 and NGC\,6946~ULX-1, which were observed with HST in the far UV range, we performed photometry of the archive data. The information on the observation date and the used cameras and filters is listed in Table~\ref{measurement}.

For the observations obtained with the Advanced Camera for Surveys (ACS), we selected the integral and distortion-corrected images in drz format. The standard initial reduction of the data from the ACS includes the procedures of bias and dark frame subtraction, flat-field correction, accounting for ``bad'' columns and ``hot'' pixels, and cosmic-hit removal. The photometry of the WFPC2 data was performed on the images in the c0f format. The masking of defective pixels, sky background subtraction and removal of cosmic hits were performed using the specialized program package {\tt HSTPHOT1.1}.

The fluxes of the optical counterparts of the \mbox{X-ray} sources were measured by means of aperture photometry using the APPHOT software package, which is based on IRAF. The aperture radii were selected in the 0\farcs15--0\farcs2 range. The main uncertainty in the point source flux measurements is due to the presence of very patchy nebulae. Following the paper by Kaaret et al.~\cite{Kaaret2010}, we estimated the contribution of the nebula to the total flux in the aperture of the object and, based on this value, calculated the point-source flux error in every filter (Table~\ref{measurement}).

\begin{table*}[th]
\setcaptionmargin{0mm} \onelinecaptionstrue
\captionstyle{normal}
\caption{HST data and results of photometry. The fluxes (given in units of ${10}^{-17}$~erg~{cm}$^{-2}$~s$^{-1}$~\AA$^{-1}$) are corrected for interstellar extinction (Table~\ref{data}). Total errors are given with allowance for the uncertainty associated with the subtraction of the contribution from the nebula. $\lambda_{\rm pivot}$---the pivot wavelength of the filter in Angstroms}
\label{measurement}
\medskip
\begin{tabular}{l|c|c|c|r@{\,$\pm$\,}l}
\hline
~~~~~~Name & \hspace{10mm}Date\hspace{10mm} & \hspace{6mm}Instrument~/~filter\hspace{6mm} & \hspace{6mm}${\lambda}_{\rm pivot}$\hspace{6mm} & \multicolumn{2}{c}{Flux} \\
\hline
NGC\,6946~ULX-1 & 2008 May 1 & ACS/SBC/F140LP & 1533 & $32.0$ & $2.7$\\
  & 2001 June 8 & WFPC2/PC1/F450W & 4557 & $2.3$ & $0.6$\\
  & 2001 June 8 & WFPC2/PC1/F555W & 5443 & $1.09$ & $0.27$\\
  & 2001 June 8 & WFPC2/PC1/F814W & 7996 & $0.27$ & $0.05$\\
\hline
Holmberg\,II~X-1  & 2006 November 27 & ACS/SBC/F165LP & 1758 & $33.0$ & $1.7$\\
  & 2009 Febrary 9 & WFPC2/WF2/F336W & 3359 & $4.62$ & $0.18$\\
  & 2007 October 3 & WFPC2/PC1/F450W & 4557 & $1.98$ & $0.18$\\
  & 2007 October 5 & WFPC2/PC1/F450W & 4557 & $1.88$ & $0.17$\\
  & 2007 October 9 & WFPC2/PC1/F450W & 4557 & $1.96$ & $0.18$\\
  & 2007 October 3 & WFPC2/PC1/F555W & 5443 & $1.06$ & $0.07$\\
  & 2007 October 5 & WFPC2/PC1/F555W & 5443 & $1.07$ & $0.07$\\
  & 2007 October 9 & WFPC2/PC1/F555W & 5443 & $1.07$ & $0.07$\\
  & 2006 December 30 & ACS/WFC/F814W & 8060 & $0.301$ & $0.012$\\
\hline
\end{tabular}
\end{table*}

The aperture corrections for the data obtained with ACS/SBC were accounted for using the $\tt{synphot}$ program package. For WFPC2 and ACS/WFC the corrections were determined by photometric measurements of three to five single stars in the apertures varying from 0\farcs15 to 0\farcs5. The so-called CTE effect may influence the results of the measurements considerably, due to the fact that when reading the CCD, the efficiency of the charge packet transfer from pixel to pixel is not equal to 100\%. When calculating the CTE corrections, we used the internet resource CTE Tool \#1\footnote{\url{http://www.stsci.edu/hst/wfpc2/software/wfpc2_cte_calc.html}}, as well as the algorithm described in section 5.1.5 of the ACS Data Handbook~\cite{Gonzaga2011}. We have applied the CTE corrections, its value does not exceed $0\fm11$.

The flux within a given filter was corrected for interstellar extinction with the assumption that the spectra of the objects are described by the power law $F_{\lambda} \propto {\lambda}^{-3}$. This procedure was performed using the calcphot command in the $\tt{synphot}$ program package.

The interstellar extinction for the objects \linebreak
 NGC\,6946~ULX-1 and Holmberg\,II~X-1 was estimated based on the $H\alpha/H\beta$ flux ratio in the nebulae. For the photoionized plasma in widely varying conditions (at temperatures of \mbox{$5000 \div 20\,000$~K} and electron densities of $10^2 \div 10^6$~{cm}$^{-3}$), $H\alpha/H\beta$ is equal to $2.86$ with an accuracy of about 5\%~\cite{Osterbrock2006}. In the case of collisional excitation this ratio may reach the value of $3.2$, and then the $H\gamma/H\beta$ ratio may be used for more reliable extinction estimates. 

We used the optical spectrum of the nebula surrounding Holmberg\,II~X-1 (taken on February~28,~2011) from the SMOKA archive~\cite{Baba2002}. To estimate the possible contribution of shocks to the observed $H\alpha/H\beta$ ratio, we analyzed the lines of the nebula. The analysis of the [S\,II]\,$\lambda\lambda6716,6734$ doublet shows that the line widths of the doublet (corrected for the instrumental profile $4.35\pm0.15$~\AA) amount to $92\pm8$~km/s. This value agrees with the results of the velocity gradient measurements for He\,II~$\lambda4686$ in~\cite{Lehmann2005}, which might indicate the presence of shock waves. However, the $H\alpha/H\beta$ and $H\gamma/H\beta$ line ratios yield values of extinction that are consistent with each other and equal to $A_V \simeq 0.23$. Therefore, the shocks contribution is fairly small. In our estimates of $A_V$ we used the extinction curve of Cardelli et al.~\cite{Cardelli1989} with $R_V=3.1$. The obtained value of extinction is in agreement with the results of~\cite{Abolmasov2007}.

For NGC\,6946~ULX-1 we have performed similar extinction estimates based on the observed fluxes in the lines of the nebula reported in~\cite{Abolmasov2007}. The $H\alpha/H\beta$ ratio results in $A_V \simeq 1.55$, whereas $H\gamma/H\beta$ yields $A_V \simeq 1.34$. Dunne et al.~\cite{Dunne2000} have discovered broad emission line components with a velocity dispersion of about $250$~km~s$^{-1}$ on the Echelle spectra of the object. It seems that in the case of the nebula around NGC\,6946~ULX-1, the shocks affect the observed line ratios. For this object, we have adopted the value $A_V \simeq 1.34$, measured from the $H\gamma/H\beta$ line ratio. Extinction corrected point-source fluxes are presented in Table~\ref{measurement}. 

The values of interstellar extinction for \linebreak
 \mbox{NGC\,1313~X-1}, \mbox{NGC\,1313~X-2} and \mbox{NGC\,5408~X-1} were taken from~\cite{Grise2008,KaaretCorbel2009,Schlegel1998} (Table~\ref{data}). In the case of \mbox{NGC\,1313~X-1}, we used the Galactic value of extinction, since there are no extinction measurements based on the Balmer decrement of the nebula for this object.

\begin{figure*}[tbp!!!]
\setcaptionmargin{5mm}
\onelinecaptionsfalse
\includegraphics[scale=0.56]{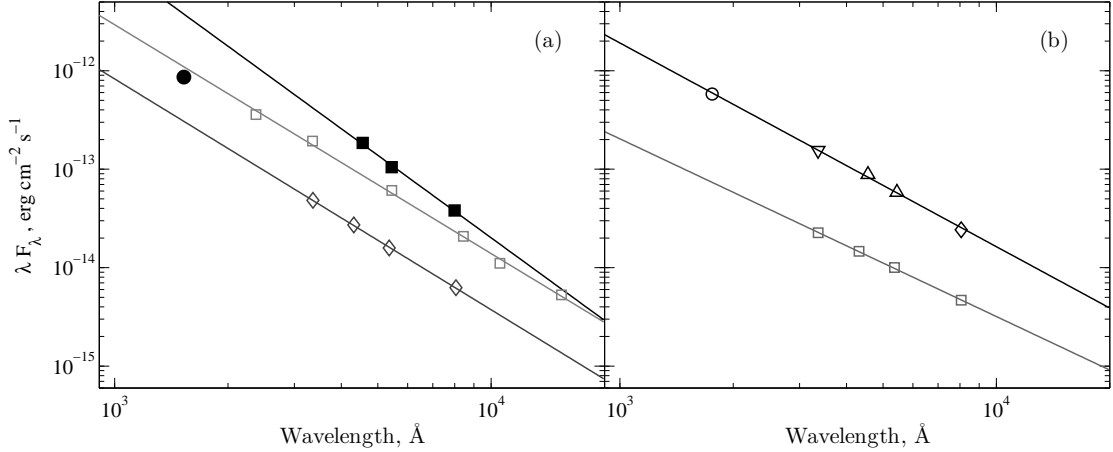}
\captionstyle{normal}
\caption{Spectral energy distributions of the ULXs. The fluxes are corrected for interstellar extinction and reduced to distance $D=3.39$~Mpc of the host galaxy of Holmberg\,II~X-1. Simultaneous data are shown by same-type symbols. The spectra are shown from top to bottom: (a) NGC\,6946~ULX-1 (filled symbols), NGC\,5408~X-1, NGC\,1313~X-2; (b) Holmberg\,II~X-1, NGC\,1313~X-1. With the exception of two objects (NGC\,6946~ULX-1 and Holmberg\,II~X-1), we show only the simultaneous data to avoid overloading the picture. The lines depict the results of the power-law ($F_\nu \propto \nu^{-\alpha}$) approximation of the fluxes.}
\label{fig:fig1}
\end{figure*}

Figure~\ref{fig:fig1} shows the spectral energy distributions for the ultra-luminous X-ray sources. The lines show the results of the $F_\nu \propto \nu^{-\alpha}$ power-law approximation of the fluxes (Table~\ref{data}). The fluxes are corrected for interstellar extinction and reduced to the Holmberg\,II~X-1 distance.

For Holmberg\,II~X-1, we used the data obtained at different epochs. The fluxes in filters F450W and F555W were averaged over three observing dates. The best approximation result is the spectral index \mbox{$\alpha = -1.07\pm0.05$}. In the case of NGC\,6946~ULX-1 we used only optical data (\mbox{$\alpha = -1.78\pm0.25$}). The ultraviolet point falls out from the power law matching the optical data. For the remaining three sources we approximated the fluxes in all available filters. Obtained $\alpha$ are given in the Table~\ref{data}. We used the 68\% confidence interval in all of these estimates.

\subsection{X-ray data}
\label{Xdata}
In this study we focus on the ULX spectra in the UV and optical ranges. Here, we take into account the X-ray data for illustrative purposes only. We therefore find it appropriate to use published results of the X-ray fits with simple models. We selected the published models that best match the observed spectra. All of these spectra have a sufficiently large total exposure time ($>10$~ks). 

The Compton scattering plays an important role in the formation of the ULX X-ray spectra at energies greater than $\sim1$~keV (e.g.,~\cite{Gladstone2009}). We will model the \mbox{X-ray} data with the Compton scattering in our next paper. Since we do not consider Comptonization here, we cut off the X-ray spectra above $1.5$~keV.

For Holmberg\,II~X-1 we selected the \linebreak
 BB+DISKPBB model with $p=0.5$ from~\cite{Kajava2012}, which was used to describe the observations from April 15, 2004. For NGC\,6946~ULX-1, there are considerably less data available in the literature than for the other objects. We took the results of the DISKBB+POWERLAW model approximation of the September 7, 2001 spectra of this source from~\cite{Berghea2008}. For the two objects in the NGC\,1313 galaxy we used the data from~\cite{FengKaaret2006}. In the case of NGC\,1313~X-1, we took the parameters of the models describing the \mbox{X-ray} spectra from October~17, 2000, January~17,~2004 and November 23, 2004. For NGC\,1313~X-2 we used the results of the above authors for December 25, 2003 and August 23, 2004.

NGC\,5408~X-1 is the only object in our list for which three sets of simultaneous observations were obtained in 2010 in the optical (HST) and X-ray (Chandra) ranges. The results of these observations are published in the paper~\cite{Grise2012}. For our study, we took the data from May 15, 2010, described by the DISKBB+POWERLAW model.

In section \ref{sec:Discusion} we separately compare the results of the approximation by our model and by DISKIR for Holmberg\,II~X-1. To this end, we re-analyzed the observations of Holmberg\,II~X-1 obtained on April 15, 2004 with the XMM-Newton PN detector (OBSID 0200470101). The reduction was carried out using the SAS 12 program package. After event filtering with FLAG\,$=0$\,PATTERN\,$\leq4$, we selected the least noise-contaminated time interval of about $14$~ks. Since the object was situated close to the CCD gap, we used the aperture with a radius of about $40\arcsec$ for extraction. The background was determined in the ring around the object. The spectrum was grouped in such a way that every bin contained at least 100 counts.

\section{The model}

\subsection{Shakura--Sunyaev supercritical disk}

The well-known paper of Shakura and Sunyaev~\cite{SS73} introduces the accretion disk model \mbox{($\alpha$-disk)}. The same paper considers the case of supercritical accretion, which occurs when the rate of matter inflow into the disk is

\begin{equation}
\dot{M}_0 > \dot{M}_{\rm Edd}=\frac{2 L_{\rm Edd}{R_{\rm in}}}{GM_{\rm BH}}=\frac{48{\pi}GM_{\rm BH}}{c\kappa},
\end{equation}  
where $L_{\rm Edd}\simeq 1.3\times10^{39}M_{10}$~erg\,s$^{-1}$ is the Eddington luminosity for a black hole with the mass $M_{\rm BH}$ expressed in units of $10$~M$_{\sun}$; \linebreak
\mbox{${\dot{M}}_{\rm Edd}=2\times 10^{19}M_{10}$~g\,s$^{-1}$} is the corresponding accretion rate; $R_{\rm in}$~is the inner radius of the accretion disk; 
$\kappa$ is the Thomson opacity. Such an accretion regime is realized in the only known superaccretor SS\,433, where ${\dot{M}}_0$ reaches the value \mbox{$\sim10^{-4}$~M$_{\sun}$/year $\sim 300\dot{M}_{\rm Edd}$} ~\cite{FabrObz,Shklovsky1981}. The supercritical properties of the disk manifest themselves below the spherization radius: 
\begin{equation}
\label{eq:Rsp}
R_{\rm sp}\sim\frac{{\dot{M}}_0\kappa}{8\pi c}.
\end{equation}
Within $R_{\rm sp}$, the disk is locally Eddington~\cite{SS73}: the force of gravity is balanced by the radiation pressure at its every point. As a result, the disk becomes geometrically thick ($H/R\sim1$). Hereafter, for certainty, we assume the half-opening angle of the funnel of the accretion disk $\theta_f$ to be equal to $45\degr$, which corresponds to $H/R=1$. 

Because of the radiation pressure, the accreting matter begins to outflow from the surface of the supercritical disk in the form of wind, which leads to a gradual decrease of the accretion rate with radius~\cite{SS73}:
\begin{equation}
\label{eq:M_R}
\dot{M}(R)=\frac{R}{R_{\rm sp}}\dot{M}_0
\end{equation}
In turn, this leads to a decrease in the amount of the released gravitational energy, and the resulting disk luminosity becomes constrained by the value 
\begin{equation}
\label{eq:Ltot}
L_{\rm tot}=L_{\rm Edd}(1+\it{a}\ln\dot{m}_0),
\end{equation}
where
\begin{equation}
\label{eq:mdot}
\dot{m}_0\equiv\frac{\dot{M}_0}{\dot{M}_{\rm Edd}}=\frac{R_{\rm sp}}{R_{\rm in}}
\end{equation}
is the dimensionless initial accretion rate, and \mbox{$\it{a}\leq1$}~\cite{Poutanen2007,SS73,Lipunova1999}. In this paper we adopt $\it{a}=1$. $L_{\rm Edd}$, is released in the standard disk above $R_{\rm sp}$, \mbox{$L_{\rm Edd} \ln \dot{m}_0$} is released in the supercritical disk.

In the case of SS\,433, the relation (\ref{eq:Ltot}) leads to the luminosity $L_{tot}\sim10^{40}$~erg\,s$^{-1}$, which corresponds to the measurements in~\cite{Murdin1980,CherepKornilov1982} and coincides with the independent estimate from~\cite{Fabrika2008W50}. The maximum of this luminosity lies in the ultraviolet part of the spectrum. In the X-ray range, where the main contribution to the observed flux is given by the emission of the cooling relativistic jets, SS\,433 is a relatively weak object ($L_X\sim10^{36}$~erg\,s$^{-1}$)~\cite{FabrObz}. 

The orientation of SS\,433 is such that we observe it close to the orbital plane and thus do not see the deep parts of the wind funnel of the supercritical disk. Initially, all of the energy release of SS\,433 originates in the accretion disk in the form of hard radiation, which then thermalized in the powerful wind of the supercritical disk~\cite{FabrObz}. For an observer who sees the funnel of the supercritical disk, SS\,433 would have an X-ray luminosity of at least $10^{40}$~erg\,s$^{-1}$. Ultra-luminous X-ray sources also demonstrate similar luminosities in the X-ray range~\cite{FengSoria2011,Swartz2004}. 

A further increase in the brightness of the supercritical disk for the observer that can see the funnel is probably due to the geometrical collimation of radiation. In the simplest case, the radiation collimation factor is determined by the relation \mbox{$B=2\pi/\Omega_f$}, where $\Omega_f$ is the solid opening angle of the supercritical-disk funnel~\cite{FabrMesch2001}. For the angle \mbox{$\theta_f\sim45\degr$}, $B\sim3$.

In their paper~\cite{SS73}, Shakura and Sunyaev have pointed out the key role of the optically thick wind outflowing from the surface of the supercritical disk in the formation of UV and optical spectra. This idea was developed further by many authors (e.g.,~\cite{Poutanen2007}).

We believe that the outflowing wind forms a funnel that has an approximately conical shape. Hereafter, we will call it ``funnel''\ or ``wind funnel''. The existence of a funnel is due to the presence of an angular momentum both in the accreting and outflowing gas, which results in the formation of a region with a low density of matter near the symmetry axis of the system. The presence of a funnel both in the supercritical disk and in the wind outflowing from it is confirmed by direct 2D RHD simulations~\cite{Ohsuga2005,Okuda2009}. 

The opening angle of the wind funnel is determined by the ratio of the Keplerian rotation velocity of the disk matter at a given radius to its outflow velocity, which is determined by the radiation pressure. The velocity of matter outflow from the surface of the supercritical disk should be close to virial~\cite{Poutanen2007,SS73}; in this case we can expect a wide funnel with the half-opening angle $\sim30\degr\div60\degr$. In this paper we adopt the opening angle of the wind funnel equal to the opening angle of the supercritical disk $\theta_f=45\degr$.

\begin{figure}[tbp!!!]
\setcaptionmargin{5mm}
\onelinecaptionsfalse
\includegraphics[scale=0.18]{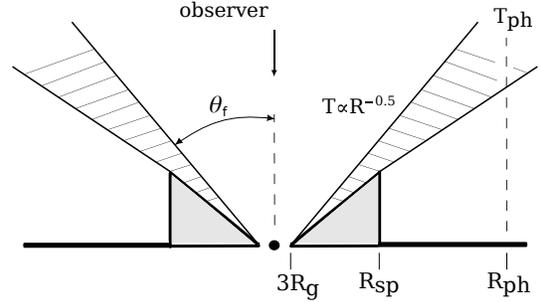}
\captionstyle{normal}
\caption{The model of a supercritical disk with a wind funnel. The figure shows the thin disk ($R > R_{\rm sp}$), the  supercritical disk ($R \leq R_{\rm sp}$), and the wind funnel constrained by the radius of the photosphere $R_{\rm ph}/\sin{\theta_f}$.}
\label{fig:fig2}
\end{figure}

Figure \ref{fig:fig2} shows the schematic model of the supercritical disk with a wind funnel . Three main components are illustrated: the thin standard disk beyond the spherization radius, $R>R_{\rm sp}$, the supercritical disk ($R_{\rm in}~{\leq}~R~{\leq}~R_{\rm sp}$), and the wind funnel, which spreads out to the radius of the photosphere ${R_{\rm ph}}/\sin{\theta_f}$. The photospheric radius confines the area of the optically thick wind; its size is determined by the geometry and velocity of the outflowing wind and by the dependence of the opacity coefficient on temperature. The exact calculation of ${R_{\rm ph}}$ is a rather difficult task; therefore, within the context of this paper, we consider it a model parameter.

The typical sizes for a Schwarzschild black hole with a mass of $10$~M$_{\sun}$ and accretion rate \mbox{$\dot{m}_0=300$} are: \mbox{$R_{\rm in}=3 R_{g}\sim10^{7}$~cm} ($R_{g}$ is the gravitational radius), \mbox{$R_{\rm sp}\sim3\times10^{9}$~cm} (formula (\ref{eq:Rsp})). The radius of the photosphere may be roughly estimated from the relation 
${R_{\rm ph}}\sim\dot{M}_w\kappa/4{\pi}V_w\cos{\theta_f}\sim \linebreak
 2\times10^{12}{(\frac{V_w}{1000~{\rm km/s}})}^{-1}$~cm (\cite{FabrikaAbolKarp2007}), where \mbox{$\dot{M}_w\sim\dot{M}_0$} is the mass loss rate, $V_w$ is the velocity of the wind at the level of the photosphere, which we assume to be  $1000$~km\,s$^{-1}$.

\subsection{Supercritical disk model (SCAD) and DISKIR model of Gierlinski et al.}

A supercritical disk can be observed as a luminous X-ray source only if the angle between the line of sight and the axis of the disk is not greater than the funnel opening angle $\theta_f$. In this case, an observer would see both the supercritical-disk radiation and the radiation of the wind funnel. In such an orientation, the central regions of the standard disk ($R\,\gtrsim\,R_{\rm sp}$), where the energy release is maximal, may be hidden from the observer by the wind. We therefore assume that the radiation of the standard disk with the luminosity $L_{\rm Edd}$ heats the wind from below. This radiation is thermalized in the wind and thus increases its energy (see below).

We assume that the radiation of the accretion disk and funnel is black-body-like. 
As was shown by Poutanen et al.~\cite{Poutanen2007}, in a supercritical disk, the behavior of effective temperature with radius is determined by the relation
\begin{equation}
\label{eq:Tsd_R}
T\propto r^{-\frac{1}{2}}
\end{equation}
where $r=R/R_{\rm in}$. Such a temperature dependence leads to a power law spectrum $F_{\nu}\propto {\nu}^{-1}$. 

The temperature at the inner radius of the disk $T_{\rm in}$ can be determined from the disk luminosity
\begin{equation}
\label{eq:L_sd_INT}
\begin{array}{c}
L = 2\int\limits_{R_{\rm in}'}^{R_{\rm sp}'} {2{\pi}R{\sigma}{T^4}(R')}\,dR' \\
= 4{\pi}{R_{\rm in}}^2{\sigma}{T_{\rm in}}^4 \int\limits_{1}^{r'_{\rm sp}} \frac{dr'}{r'\sin{\theta_f}}~.
\end{array}
\end{equation}
Hereinafter, by $R'=R/\sin{\theta_f}$ we understand the distances measured along the disk and wind funnel surface; $r'~=~R'/R_{\rm in}~=~r/\sin{\theta_f}$. The luminosity of the supercritical disk ($R\,\leq\,R_{\rm sp}$), the mass of the black hole, and the initial accretion rate are bound by the relation  $L_{\rm bol}=\it{a} L_{\rm Edd} \ln{\dot{m}_0}$, $\it a$=1 (\ref{eq:Ltot}). 
Using (\ref{eq:L_sd_INT}) we find the temperature
\begin{equation}
T_{\rm in} = {\left[ \frac{L_{\rm Edd}}{4{\pi}{\sigma}{{R_{\rm in}}^2}}{{~\sin{\theta_f}}}\right]}^{\frac{1}{4}} .
\end{equation}

A noticeable portion of the supercritical-disk radiation may be absorbed by the wind, followed by re-emission at lower frequencies. In this respect, the SCAD model is similar to the model proposed by Gierlinski et al. (2009)~\cite{Gierlinski2009}, DISKIR, which is currently used to explain the spectral energy distributions of ULXs~\cite{TaoHoII2012,Tao2011,Grise2012}. 

DISKIR is based on the standard ``multi-temperature disk''\ model (DISKBB), but takes into account the effects of Comptonization and irradiation of the disk by hard photons. The inner disk is covered by a semi-transparent gas cloud (``hot flow'') up to the radius $r_{\rm irr}$. The hot gas heats the outer parts of the accretion disk, which results in a considerable luminosity in the UV-optical range. Initially, this model has been applied to the low-mass X-ray source XTE J1817-330~\cite{Gierlinski2009}.

The DISKIR model contains 8 parameters~\cite{Gierlinski2009}. The observed UV, optical and near-infrared fluxes determine the parameters $f_{\rm out}$ (fraction of reprocessed radiation) and $r_{\rm out}$ (outer radius of the accretion disk). The X-ray part of the spectrum constrains the other six parameters, including $kT_{\rm in}$, $\Gamma$ (power law photon index for the Comptonized radiation), $kT_{\rm e}$ (temperature of the Comptonizing electrons), and  $L_{\rm c}/L_{\rm d}$ (ratio of the Compton tail luminosity to the luminosity of the accretion disk), which determine the shape of the spectrum in the inner regions of the disk. The parameters $r_{\rm irr}$ (radius of the inner-disk area covered with ``hot flow'') and $f_{\rm in}$ (fraction of the Comptonized radiation that ends up in the inner disk) are responsible for the reprocessing of radiation in the $r{\leq}r_{\rm irr}$ region. With the exception of the $L_{\rm c}/L_{\rm d}$ ratio, the last 6 parameters have almost no effect on the simulated optical fluxes.

For the temperature $T$ of the outer disk heated by the radiation from the central parts, Gierlinski et al. adopt $T{\propto}r^{-\frac{1}{2}}$. In the case of heating of the standard (flat) accretion disk by the central source, the temperature of the irradiated surface of the disk will be determined by $T \propto r^{-\frac{3}{4}}$, the same as at gravitational energy release. Gierlinski et al. suggested that the disk may be warped; in this case, the temperature may depend on radius as $T {\propto} r^{-\frac{1}{2}}$. Since the temperature in the standard disk is determined by the relation $T \propto r^{-\frac{3}{4}}$, the reprocessed radiation should dominate over the gravitational energy released in the disk at large radii.

\subsection{The SCAD model parameters}

In our model we assume that the relation $T{\propto} {r}^{-\frac{1}{2}}$ also applies for the wind funnel. Such a dependence may be caused by the geometry of the wind (warped wind) or the presence of scattering matter filling the funnel. 

Similar to~\cite{Gierlinski2009}, we determine the re-emitted flux by the following equation:
\begin{equation}
F = \frac{f_{\rm out}L_{\rm tot}}{4{\pi}{R'}^2},
\end{equation}
where $f_{\rm out}$ is the fraction of the bolometric flux reprocessed in the wind at the radius $R'$, and $L_{\rm tot}$ is the bolometric luminosity (\ref{eq:Ltot}). The resulting spectrum of the wind funnel, as well as the spectrum of the supercritical disk, will be power law, $F_\nu \propto \nu^{-1}$. 

The parameter $f_{\rm out}$ is a combination of many geometrical and physical characteristics of the wind, such as the fraction of the accretion-disk radiation absorbed by the wind, the albedo, and the absorbed-radiation thermalization efficiency. The supercritical disk with luminosity $L_{\rm Edd}\ln{\dot{m}_0}$ heats the wind from the side of the funnel, and the standard disk at \mbox{$R > R_{\rm sp}$} with luminosity $L_{\rm Edd}$ heats the same wind from below. 

As a result, the total flux of the supercritical disk with a wind funnel will be described by
\[ F_{\rm tot} = \sigma{T^4}_{\rm in} \left\{ \begin{array}{ll}
{({r'}\sin{\theta_f})}^{-2},~~1~\leq~r'~\leq~{r_{\rm sp}}'\\
{\frac{f_{\rm out}}{\sin{\theta_f}}(1+\ln{\dot{m}_0}){r'}^{-2}}, \\
{r_{\rm sp}}'~\leq~r'~\leq~{r_{\rm ph}}'\\
\end{array} \right. \]

Thus, our model has seven parameters. The four main parameters that fully determine the shape of the model spectra are: black hole mass $M_{\rm BH}$ \mbox{($L_{\rm Edd} \propto M_{\rm BH}$)}, initial normalized accretion rate $\dot{m}_0$, fraction of bolometric flux reprocessed in the wind $f_{\rm out}$, and wind photosphere radius $R'_{\rm ph}$. Next two parameters are the opening angles of the supercritical disk and wind funnel, and in this paper, we assume them to be equal, $\theta_f=45\degr$. We set the seventh parameter $a$ (\ref{eq:Ltot}) equal to one, assuming within this paper that advection is absent in the accretion disk (see section \ref{sec:Discusion}).   

\begin{figure}[tbp!!!]
\setcaptionmargin{5mm}
\onelinecaptionsfalse
\includegraphics[scale=0.5]{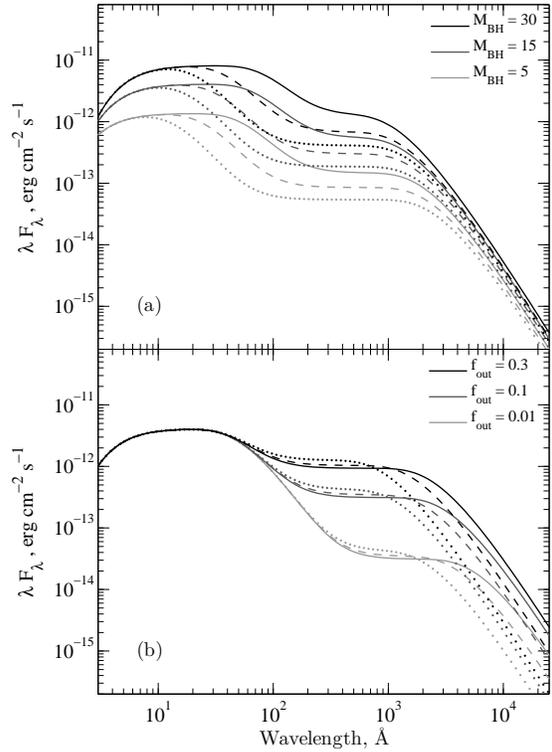}
\captionstyle{normal}
\caption{Model spectral energy distributions (SEDs)  for  the  supercritical  disk  with  a  wind funnel.   Different  grey  tones  correspond  to different black hole masses $M_{\rm BH}$ for the accretion  rates $\dot{m}_0 =$ 30 (dots lines), 100 (dashed lines)  and  500  (solid lines) (a),  and different values of  $f_{\rm out}$  for  the  photospheric  radii \mbox{$R'_{\rm ph} / R_{\rm in} =$ $0.5 \times 10^{5}$} (dots lines), $1.5\times 10^{5}$ (dashed lines) and $3\times 10^{5}$ (solid lines)~(b). The values of the fixed parameters are: \mbox{$f_{\rm out}= 0.1$} and $R'_{\rm ph}=10^{12}$~cm ($a$), $M_{\rm BH}=15$~M$_{\sun}$ and $\dot{m}_0=200$~($b$).}
\label{fig:fig3}
\end{figure}

Figure \ref{fig:fig3} shows the model spectral energy distributions. The fluxes were computed for the distance D~=~3~Mpc. Two flat areas on the spectra correspond to the power law $F_\nu \propto \nu^{-1}$. The radiation of the supercritical disk dominates in the X-ray range. In the UV/optical range the observed spectra are mainly determined by the radiation of the irradiated wind. In the long-wavelength part, the flux is determined by the Rayleigh-Jeans spectrum of the wind funnel, $F_\nu \propto \nu^{2}$.

Figure \ref{fig:fig3}a demonstrates the change in the spectra as a function of black hole mass $M_{\rm BH}$ and accretion rate $\dot{m}_0$. With increasing mass, the Eddington limit $L_{\rm Edd}$, and therefore, the bolometric luminosity of the accretion disk proportionally increase. The luminosity of the wind funnel also linearly grows with $M_{\rm BH}$ when $f_{\rm out}=const$. Therefore, on the whole, the radiation spectrum shifts to the domain of higher fluxes.

An increase in the accretion rate at constant black hole mass leads to the same change in the geometric size of the supercritical disk and a decrease in the gas and radiation temperature in its outer regions. For the temperature at the spherization radius, the following relation is valid: $T_{\rm sp} \propto {\dot{m}_0}^{-\frac{1}{2}}$. The temperature decrease in the outer parts of the disk is due to the increase of the radiating area (${R_{\rm sp}}^2 {\propto} {\dot{m}_0}^2$). As a result, the flat part of the X-ray spectrum extends to the long-wavelength region (Fig.~\ref{fig:fig3}a).

The dependence of the spectra on $f_{\rm out}$ and $R'_{\rm ph}$ at constant values of $M_{\rm BH}$ and $\dot{m}_0$ is shown in Fig.~\ref{fig:fig3}b. The parameter $f_{\rm out}$ mainly acts as a scale factor for the UV and optical parts of the spectrum, since the flux linearly grows with the fraction of absorbed and re-emitted energy. At the same time, the dependence of radiation temperature on the amount of energy reprocessed by the wind is fairly weak and proportional to ${f_{\rm out}}^{\frac{1}{4}}$. For this reason, the wind temperature decreases with decreasing $f_{\rm out}$, and the flat part of the UV-optical spectrum shifts slightly to the long-wavelength region (Fig.~\ref{fig:fig3}b). 

The radius of the photosphere determines the outer boundary of the blackbody emitting wind. The increase of the wind-funnel size at a constant irradiating flux leads to the decrease in the intensity and temperature of the wind radiation. Thus, the increase of the photospheric radius of the wind funnel extends the flat part of the UV spectrum into the optical range.

\section{Results}

We have fitted the ULX SEDs with the SCAD model. As X-ray data, we used the published models (including absorption models) that best describe the observed spectra (section~\ref{Xdata}). Based on the parameters of these models, we have constructed the absorbed spectra. We used models instead of the observed spectra because the data approximation was carried out without {\tt XSPEC}.

When approximating the absorbed X-ray spectra with our model, we determined the new value of $N_H$. We used the extinction model from~\cite{wabs1983}. In the optical range we used the extinction-corrected fluxes. Unlike in the X-ray range, the optical interstellar extinction is usually well-known from the spectroscopy of the nebulae surrounding the objects.

Using the relation $A_V= 4.5\times 10^{-22} N_H$~\cite{Gorenstein1975extinct}, we have compared the extinctions in the optical and \mbox{X-ray} ranges (Tables~\ref{data} and \ref{param}). In all cases, the value of $A_V$ obtained from the optical data is smaller than the $A_V$ obtained from the X-ray spectra. The difference ${(A_V)}_{\rm X-ray} - {(A_V)}_{\rm opt}$ is \mbox{$\sim 0\fm2 \div 0\fm8$}, the average value is $\sim 0\fm5$.

Table~\ref{param} shows the results of spectral energy distribution simulations for all of the five objects. The table includes the values of column density taken from literature ${(N_H)}_{\rm lit}$ and the ones obtained by us, $N_H$, inner radius temperatures, $k T_{\rm in}$,  black hole masses $M_{\rm BH}$, normalized accretion rates $\dot{m}_0$, fractions $f_{\rm out}$ of the bolometric flux of the accretion disk reprocessed in the wind, and radii of the wind-funnel photosphere $R'_{\rm ph}$. We do not quote the parameter errors, because in our approximations we used X-ray spectra, which are not direct observational data. The obtained values of $M_{\rm BH}$ correspond to stellar-mass black holes, and $\dot{m}_0$ are comparable to the accretion rate in the SS\,433 system. The average values of parameters $f_{\rm out}$ and $R'_{\rm ph}$ for the five studied objects are $2.8\times 10^{-2}$ and $2.3\times 10^{12}$~cm, respectively.

\begin{table*}[tbp]
\setcaptionmargin{0mm} \onelinecaptionsfalse
\captionstyle{normal}
\caption{Model parameters of ULXs. ${(N_H)}_{lit}$---column density taken from literature, $N_H$---column density obtained from our fit to the X-ray spectra; $k T_{\rm in}$---inner disk temperature; $M_{\rm BH}$---black hole mass; $\dot{m}_0$---initial normalized accretion rate; $f_{\rm out}$---fraction of the bolometric flux reprocessed in the wind funnel; $R'_{\rm ph}$---radius of the photosphere}
\label{param}
\medskip
\begin{tabular}{l|c|c|c|c|c|c|c}
\hline
~~~~~~~~Name &  ${(N_H)}_{\rm lit}$ &  $N_H $ & $k T_{\rm in}$ & $\hspace{5mm}M_{\rm BH}\hspace{5mm}$ & $\hspace{5mm}\dot{m}_0\hspace{5mm}$ & $\hspace{5mm}f_{\rm out}\hspace{5mm}$ &  $R'_{\rm ph}$ \\
 & $(\times\,10^{21}~$cm$^{-2})$ & $(\times\,10^{21}~$cm$^{-2})$ & (keV)  & $(M_{\sun})$ &  & ($\times\,10^{-2}$) &  ($\times\,10^{12}$~cm)\\
\hline
Holmberg\,II X-1 & 0.87$^1$ & 0.94 & 0.81 & 17  & 270 & 4.1 & 2.3\\
NGC\,6946 ULX-1 & 6.12$^2$ & 4.3   & 0.77 &  20 & 180 & 4.2 & 3.5\\
NGC\,1313 X-1 & 2.5$^3$ & 2.5      & 0.82 & 16  & 150 & 0.18 & 1.6\\
NGC\,1313X-2 & 2.7$^3$ & 2.5       & 0.99 & 8   & 200 & 1.8 & 1.3\\
NGC\,5408 X-1 & 1.27$^4$ & 0.97    & 0.85 & 14  & 250 & 3.8 & 2.6\\
\hline
\end{tabular}
\begin{flushleft}
\footnotesize{$^1$Kajava et al. 2012~\cite{Kajava2012}, $^2$Berghea et al. 2008~\cite{Berghea2008}, $^3$Feng\,\&\,Kaaret 2006~\cite{FengKaaret2006}, $^4$Grise et al. 2012~\cite{Grise2012}.}
\end{flushleft}
\end{table*}

\begin{figure*}[tbp!!!]
\setcaptionmargin{5mm}
\onelinecaptionsfalse
\includegraphics[scale=0.56]{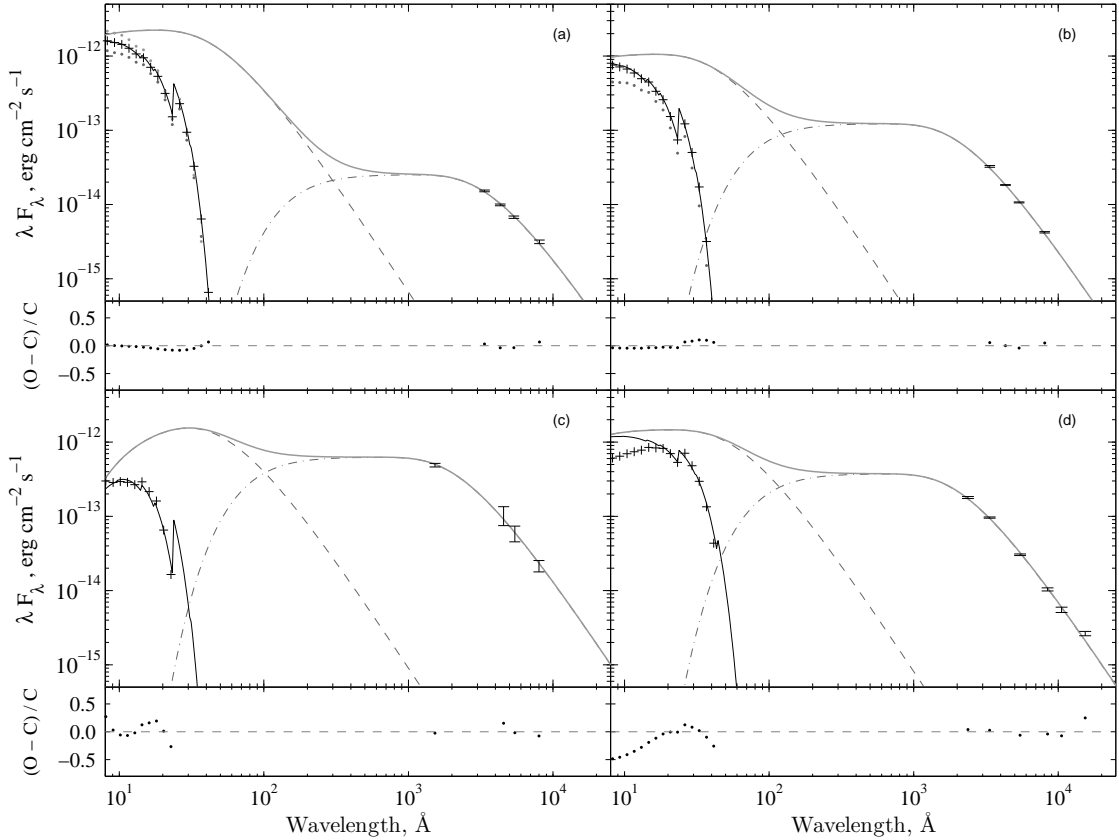}
\captionstyle{normal}
\caption{The observed energy distributions of ULXs and their fit with the model of a supercritical  disk with a wind funnel for NGC\,1313~X-1 (a), NGC\,1313~X-2 (b), NGC\,6946~ULX-1 (c) and NGC\,5408~X-1 (d). The grey dashed curve shows the spectrum formed in the supercritical accretion disk, the  grey dash-dotted  curve  shows  the  spectrum  formed  in  the  wind funnel, and  the  grey solid  curve  is  the integral  spectrum.  The black crosses (and grey dots in the cases of  NGC\,1313~X-1 and  NGC\,1313~X-2) represent the absorbed X-ray spectra, and the black curve shows our model approximation of the X-ray spectrum. The lower panels show the relative deviations of the observed energy distributions from the simulated ones.}
\label{fig:fig4}
\end{figure*}

\begin{figure}[tbp!!!]
\setcaptionmargin{5mm}
\onelinecaptionsfalse
\includegraphics[scale=0.5]{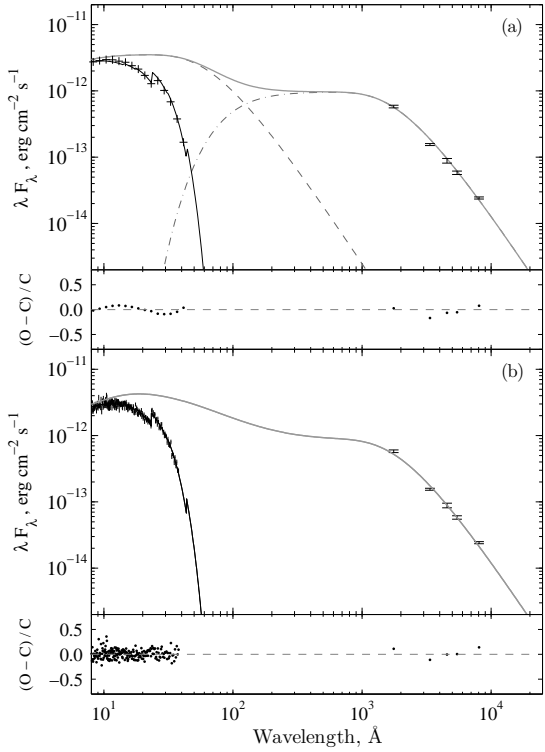}
\captionstyle{normal}
\caption{Comparison of the SCAD (a) and DISKIR (b) models for Holmberg\,II~X-1. Designations as in Fig.~\ref{fig:fig4}.}
\label{fig:fig5}
\end{figure}

Figures~\ref{fig:fig4} and \ref{fig:fig5}a show the observed spectral energy distributions and their approximation by the SCAD model. The optical fluxes corrected for interstellar extinction are shown with the error bars. The black crosses and grey dots indicate the absorbed X-ray spectra reconstructed from literature. To avoid overloading the images, the spectra demonstrating object variability in the X-ray range are shown only for NGC\,1313~X-1 and NGC\,1313~X-2 (grey dots).

The curves show the model and its components in different spectral regions. The spectra forming in the supercritical accretion disk (dashed curve) and wind funnel (dot-dashed curve) together give the integral spectrum, shown by upper curve. At higher energies the solid line shows our model approximation of the X-ray data. The lower panels show the discrepancy between the observations (O) and the model (C). 

{\emph{NGC\,1313~X-1, \mbox{NGC\,1313~X-2}}}.  
For \mbox{NGC\,1313~X-1} and NGC\,1313~X-2 the model SEDs describe both the optical and X-ray data well (Fig.~\ref{fig:fig4}a,b). The deviations do not exceed 10--15\%. A distinctive feature of NGC\,1313~X-1 is the very luminous (compared to the optical fluxes) X-ray spectrum, which is unusual for our sample. The value of $f_{\rm out}$ for this object is more than one order of magnitude less than the average value in other objects. 
{\sloppy

}

{\emph{Holmberg\,II~X-1}}.
The model describes the energy distribution of Holmberg\,II~X-1 fairly well (Fig.~\ref{fig:fig5}a). The only outlier is the flux in filter F336W ($\lambda_{\rm pivot}$\,=\,3359~\AA) that approximately corresponds to filter U, which may be due to a Balmer jump or object variability. Note that non-simultaneous optical observations were used in the approximation of this ULX.

{\emph{NGC\,6946~ULX-1}}. 
NGC\,6946~ULX-1 (Fig.~\ref{fig:fig4}c) demonstrates a very low X-ray flux (both in the soft and hard ranges) compared to the UV and optical fluxes, which distinguishes this object from the other ULXs in our sample. To reconcile the model with the X-ray data, we excluded the areas with $R \leq 5R_{\rm in}$. We suggest that this object might be visible at an angle slightly larger than the opening angle of the wind funnel $\theta_f$. In this case, the innermost parts of the supercritical-disk funnel may be hidden from the observer by the outer parts of the wind.

{\emph{NGC\,5408~X-1}}.
For NGC\,5408~X-1 (Fig.~\ref{fig:fig4}d), we were unable to achieve a good agreement between the model and the X-ray data at energies higher than $\sim0.8$~keV ($\sim16$~\AA), where the simulated fluxes are twice as high as the observed ones. In our next paper we intend to include the Comptonization in the inner parts of the supercritical accretion disk. Hot wind may cover the inner parts of the disk, absorbing and Comptonizing their radiation. We suppose that in the case of NGC\,5408~X-1, this wind captures substantially more disk radiation than in the other objects. We attribute the discrepancy between the model and the X-ray data to the presence of such a wind. In the optical range, our model reconstructs the observed fluxes fairly well. A slight excess is observed only in the IR range in filter F160W ($\lambda_{\rm pivot}$\,=\,15369~\AA).

Thus, on the whole, the SCAD model gives good agreement between the observed and simulated energy distributions of ULXs in the wavelength interval from soft X-ray to optical.

Figure \ref{fig:fig5} shows the SED approximation by the SCAD model ($a$) and by DISKIR ($b$) for Holmberg\,II~X-1. To correctly compare the approximation results of the two models, we used the same spectra from April 15, 2004. In Fig.~\ref{fig:fig5}a, this X-ray spectrum is shown as the best model from~\cite{Kajava2012}, and Fig.~\ref{fig:fig5}b shows the spectrum processed by us (section~\ref{Xdata}). When using DISKIR, the extinction in the X-ray region was reconstructed using the WABS model, and in the optical range, we approximated the extinction-corrected fluxes.

Following~\cite{TaoHoII2012,Gierlinski2009}, we fixed the parameters of electron temperature, \mbox{$kT_{\rm e}=100$~keV}, inner-disk radius, \mbox{$r_{\rm irr}=1.1$}, and fraction of Comptonized radiation thermalized in the \mbox{$r<r_{\rm irr}$} region, \mbox{$f_{\rm in}=0.1$}. The remaining parameters were considered free. The best agreement between the model and observations with \mbox{$\chi^2/dof=340.0/299$} was obtained with the \mbox{values}
\begin{list}{}{
\setlength\leftmargin{0in}
\setlength\topsep{0in}
\setlength\parsep{0in}
\setlength\itemsep{0in}
}
\item \mbox{$N_H=(1.06\pm0.16)\times 10^{21}$~cm$^{-2}$}, \\
\item \mbox{$kT_{\rm in}=0.12\pm0.03$~keV},   \mbox{$\Gamma=2.53\pm0.03$}, \\
\item  \mbox{$L_{\rm c}/L_{\rm d}=2.7\pm1.7$},  \mbox{$f_{\rm out}=(3.3\pm0.5)\times10^{-2}$}, \\
\item \mbox{$\log{(r_{\rm out})}=3.23\pm0.12$}~~ \mbox{(in $R_{\rm in}$ radii),} \\
\item \mbox{$norm=(7\pm3)\times10^2$}. 
\end{list}

Both models describe the X-ray and optical data equally well. The most significant difference between the model energy distributions is in the slope of the X-ray spectrum in the $\sim0.3$--$1$~keV region \mbox{($12$--$40$~\AA)}, where the model of the supercritical disk predicts lower (compared to DISKIR) fluxes and a flat spectrum. Both models result in similar values of $N_H$, however, the value of this parameter in DISKIR is larger by $\sim$10\%. This is caused by the necessity to compensate for the higher (compared to the SCAD model) fluxes in the soft X-ray range. 

Nevertheless, the models predict vastly different temperatures at the inner disk radius. For Holmberg\,II~X-1, the SCAD model gives \mbox{$kT_{\rm in}\simeq0.8$~keV}, whereas DISKIR yields an estimate of \mbox{$kT_{\rm in}\simeq0.1$~keV}. In turn, the low temperature at the inner radius leads to a black hole mass increase. Using normalization in the DISKIR model and a mean cosine of the angle between the disk and the line of sight \mbox{$<\cos{\theta}>=0.5$}, we derive \mbox{$R_{\rm in}\sim10^9$~cm}, and the corresponding mass \mbox{$M_{\rm BH}\sim1000~$M$_{\sun}$} \mbox{($L/L_{\rm Edd} \sim 0.05$)}. The SCAD model gives \mbox{$M_{\rm BH}\simeq17~$M$_{\sun}$} for Holmberg\,II~X-1.

\section{Discussion}
\label{sec:Discusion}

A comparison of our model and DISKIR in the case of Holmberg\,II~X-1 (Fig.~\ref{fig:fig5}) shows that both models well describe the observed optical to soft X-ray spectrum. The DISKIR model has already been applied to Holmberg\,II\,X-1 in the study~\cite{Tao2011}. Despite the fact that different observational data was used by these authors and us (Fig.~\ref{fig:fig5}b), the parameters of the best SED model in both cases turned out approximately the same. In particular, for Holmberg\,II~X-1,~\cite{Tao2011} derived the masses 400 and 600~M$_{\sun}$ from the obtained inner-radius temperature \mbox{$kT_{\rm in} = 0.09$--$0.24$~keV}.

For the other ULXs to which the DISKIR model has been applied, NGC\,6946~ULX-1~\cite{BergheaMF162012} and NGC\,5408~X-1~\cite{Grise2012}, the authors obtained similar  $kT_{\rm in}$ temperatures, $0.12$ and $0.13$~keV, respectively. Obviously, all of these temperatures correspond to the IMBH mass range; the~\cite{Gierlinski2009} model gives large black hole masses for the ULXs. Indeed, the DISKIR model presumes a standard accretion disk, where the huge X-ray luminosities of ULXs can be sustained only in the case of IMBHs. However, the recent data~\cite{Poutanen2012} reliably shows that ULXs are stellar mass black holes, the progenitors of which are massive stars with masses of 50--100~M$_{\sun}$. Our SED model, which is based on the SCAD approximation of Shakura and Sunyaev, presumes black hole masses of 10--20~M$_{\sun}$, which can well be achieved in the process of stellar evolution. Therefore, this model is compatible with the new data on the nature of ULXs.

The shape of the spectrum in the soft X-ray region may probably serve as a critical test for distinguishing between the SCAD and DISKIR models. In this region, $12$--$40$~\AA\ ($\sim 0.3-1$~keV), the SCAD model demonstrates a flat spectrum ($\nu F_{\nu}$), whereas in the DISKIR model there cannot be any flat parts anywhere, as it is based on the standard $\alpha$-disk. In DISKIR, when $L_{\rm c}/L_{\rm d}$ changes from $0.3$ to $7$, the maximum in the model spectrum shifts from $0.3$~keV energies to $1.0$~keV.

In Figure \ref{fig:fig5}, the differences between the SCAD and DISKIR models are not very conspicuous, both the best models appear the same. This is due to the fact that the other parameters of the models, such as $N_H$ and disk luminosity, when combined in a certain way, can smooth out the differences between the models. Nevertheless, the absence of a flat spectrum in the DISKIR model and the presence of such in SCAD are essential. High signal-to-noise spectra may probably be used to differentiate between the two models.

In this paper we assumed some of the model parameters to be constant, e.g., the opening angle of the supercritical-disk funnel $\theta_f$ and the opening angle of the wind funnel, which is formed by the wind of the supercritical disk, were set equal to $45\degr$. Within the framework of our simple approach, a change in the opening angle of the funnel will change the observed luminosity, and therefore, as a result, will somewhat re-determine the black hole mass, \mbox{$M_{\rm BH} \propto {\sin{\theta_f}}^{-2}$}. This equally applies to the parameter $a$ (\ref{eq:Ltot}), which we set equal to one, as would be the case in the absence of any advection in the accretion disk~\cite{SS73}. Here \mbox{$M_{\rm BH} \propto (1 + a \ln{\dot{m}_0})^{-1}$}. At \mbox{$\dot{m}_0 = 300$}, a decrease of $a$ from 1 to the expected value $\sim0.5$ (\cite{MineshigeAdvec2000,WangAdvec2013}) will lead to a 1.75 times increase of the mass estimate. In this paper we also disregard the collimation of radiation by the funnel in the supercritical disk and wind, which can amount to $\sim2$--$3$ if the funnel opening angle is $45$--$55\degr$. Taking into account the collimation of the radiation toward the observer will lead to a decrease in the black hole mass estimate by about $\sim$2 times.
 
In the current version of the model we disregard Comptonization---a primary component in the formation of X-ray spectra. We assume that the inner parts of the supercritical disk are covered by a semi-transparent hot gas. By analogy with SS\,433, one can claim that such gas should also exist in the funnel. This gas may not become a part of the jets during their formation and collimation, or, it is this gas that constitutes the relativistic jets of SS\,433 at the moment of their formation. The density of this gas may roughly vary according to the relation $n \propto r^{-2}$; this gas should leave the funnel. The presence of such gas will lead to the Comptonization of the radiation of the supercritical-disk inner parts, which are covered by this gas. 

Medvedev and Fabrika~\cite{Medvedev2010} have discovered an \mbox{X-ray} emission component in the XMM data of SS\,433 that has a flat spectrum. The authors suggested that this is Comptonized radiation, formed in the inner parts of the funnel and reflected by the funnel outer walls. An analogy with SS\,433 gives us a reason to assume that the inner regions of supercritical disks may contain hot outflowing gas. After including Comptonization, which we plan to do in the next paper, we will be able to approximate the ULX spectra in the entire standard X-ray range.

In this paper we assumed that the UV and optical radiation in ULXs is formed in the accretion disks (standard, in the case of IMBHs, or supercritical in the case of stellar-mass black holes). However, this radiation may be formed at the surface of the donor. The optical spectra of ULX counterparts show that all the objects ever observed have photosphere temperatures $T \gtrsim 30\,000$~K, and it is unlikely that we observe the donor's own radiation in the UV and optical ranges~\cite{Fabrika2013}. 

Let us assume that the IMBH heats the donor's surface and the latter overfills its critical Roche lobe. The size of the donor in units of orbital separation~\cite{Paczynski1971} is \mbox{$r_{\rm D}/a \sim 0.46  q^{1/3}$}, where \mbox{$q = M_{\rm D}/M_{\rm BH}$} and  $M_{\rm D}$ the donor mass. If we adopt the X-ray isotropical luminosity of ULXs \mbox{$L_x = 10^{40}  L_{40}$~erg/s}, and the donor surface albedo $A = 0.5$, we derive the bolometric luminosity of the heated donor \mbox{$L_{\rm bol} \sim A L_x (r_D/a)^2/4 \sim 1.2 \times 10^{37} L_{40}  q_{0.01}^{2/3}$~erg/s}, where $q_{0.01}$ the mass ratio in units 0.01. This luminosity corresponds to the bolometric magnitude of the donor $M_{\rm bol} \approx -4.0$. For a star with the temperature, $T \gtrsim 30\,000$~K the bolometric correction is $BC \lesssim -3.0$. We find the absolute magnitude of the heated donor $M_{V} \gtrsim -1$. This is substantially smaller than the counterpart luminosity observed in ULXs (${\rm M_V} \approx - 6 \pm 1$~mag~\cite{Tao2011}). Therefore, the scenario of the formation of optical and UV spectra on the irradiated donor surface must be excluded.

\section{Conclusions}

In this paper we have described a model of energy distribution in supercritical accretions disks, which is based on the Shakura--Sunyaev (1973) supercritical disk conception. We applied this model to ultra-luminous X-ray sources. In this model, the disk becomes thick at distances from the center less than the disk spherization radius, $H/R \sim 1$; the temperature depends on radius as $T \propto r^{-1/2}$. Here, the disk luminosity is $L_{\rm bol} \simeq L_{\rm Edd}  \ln{\dot{m}_0}$, and also, in this region, the powerful wind that forms a funnel above the disk emerges. At distances greater than the spherization radius, the disk is thin; its total luminosity is equal to the Eddington luminosity $L_{\rm Edd}$.

We believe that for an observer that can see the entire funnel of the supercritical disk (who is situated close to the accretion-disk axis, $< \theta_f$), the thin accretion disk at the radii $r \gtrsim r_{\rm sp}$ is invisible, as it is hidden by the extended wind of the supercritical disk. The thin disk irradiates this wind from below. From the side of funnel, the wind is irradiated by the supercritical disk. In this paper we adopt the following dependence of the irradiated wind temperature on distance: $T \propto r^{-1/2}$; this dependence may be due to both the curvature of the wind, and the presence of gas in the funnel, in which jets possibly spread. Both these effects cause an increase in the efficiency of irradiation of the wind funnel walls.

In this paper we do not consider the Comptonization of the radiation emerging from the central regions of the supercritical disk funnel. Accounting for Comptonization is absolutely necessary in this model, but we restrict this study to the analysis of the spectra from $\sim 1.5$~keV to the lower energies. By analogy with SS\,433, we assume that there should exist a hot wind in the inner parts of the funnel; this is the gas of the forming jets (and also the gas that contributes to the formation of the jets, but does not become part of them). This wind covers the inner parts of the supercritical disk and Comptonizes the radiation from these parts. 

The described supercritical disk model is technically similar to the DISKIR model~\cite{Gierlinski2009}, which was initially developed for LMXBs, but has already been used for the analysis of ULX SEDs. The models differ in disk type (standard---supercritical) and in the fact that in one model, the disk is irradiated at the distances $r > r_{\rm irr}$, and in the other, the wind is irradiated at $r > r_{\rm sp}$. We believe that distinguishing between the models is possible in the soft X-ray range ($\sim 0.3-1$~keV), since the SCAD model has a flat region in the spectrum, whereas DISKIR cannot have flat regions anywhere, as it is based on the standard $\alpha$-disk. Since the differences between the models can be masked by varying the parameters (e.g., $N_H$), obtaining X-ray spectra with high signal-to-noise ratios is a necessary condition for distinguishing between the models.

We see a rather important difference between the models in the fact that DISKIR, when applied to ULXs, explains their energy distributions only in the case of black holes with masses of several hundred solar masses (IMBHs), whereas the SCAD model offers fairly adequate black hole masses $\sim10$~M$_{\sun}$, which can be obtained in the course of evolution of massive stars. Recent results of ULX studies favor the nature of these objects as supercritical accretion disks with stellar-mass black holes, but not IMBHs.

To adequately describe the ULX SEDs in any model, in particular, in SCAD, one needs to obtain simultaneous observational data in the X-ray to optical range, and most importantly, in the soft X-ray and UV spectral regions. Since in both the SCAD and DISKIR models the UV and optical spectra are formed as a result of irradiation of the outer regions by the radiation from the inner regions, the observations must be simultaneous in all of these ranges.

\begin{acknowledgments}
The authors thank O.~Galazutdinova and A.~Valeev for their help with the HST data reduction. Our results are based on observations made with the NASA/ESA Hubble Space Telescope, obtained from the data archive at the Space Telescope Science Institute. STScI is operated by the Association of Universities for Research in Astronomy, Inc. under NASA contract NAS 5-26555. Our results are also based in part on the data collected with the Subaru Telescope and obtained from SMOKA, which is operated by the Astronomy Data Center, National Astronomical Observatory of Japan. This work was supported by the RFBR grant N\,10--02--00463, the Leading Scientific School grant N\,4308.2012.2 and the Ministry of Education and Science of the Russian Federation, projects N\,8406 and N\,8416.
\end{acknowledgments}

\end{document}